\documentclass[twocolumn,showpacs,aps,prl,superscriptaddress]{revtex4}

\usepackage{graphicx}
\usepackage{dcolumn}
\usepackage{amsmath}
\usepackage{epsfig}

\RequirePackage{xspace}

\usepackage{relsize}
\def\babar{\mbox{\slshape B\kern-0.1em{\smaller A}\kern-0.1em
    B\kern-0.1em{\smaller A\kern-0.2em R}}}

\def\Kbar  {\kern 0.2em\overline{\kern -0.2em K}{}\xspace}

\def\Kz    {\ensuremath{K^0}\xspace}
\def\Kzb   {\ensuremath{\Kbar^0}\xspace}
\def\KzKzb {\ensuremath{\Kz \kern -0.16em \Kzb}\xspace}
\def\Kp    {\ensuremath{K^+}\xspace}
\def\Km    {\ensuremath{K^-}\xspace}

\def\KpKm  {\ensuremath{\Kp \kern -0.16em \Km}\xspace}

\def\Dbar    {\kern 0.2em\overline{\kern -0.2em D}{}\xspace}

\def\Dz      {\ensuremath{D^0}\xspace}
\def\Dzb     {\ensuremath{\Dbar^0}\xspace}
\def\DzDzb   {\ensuremath{\Dz {\kern -0.16em \Dzb}}\xspace}
\def\Dp      {\ensuremath{D^+}\xspace}
\def\Dm      {\ensuremath{D^-}\xspace}

\def\DpDm    {\ensuremath{\Dp {\kern -0.16em \Dm}}\xspace}

\def\B       {\ensuremath{B}\xspace}
\def\Bbar    {\kern 0.18em\overline{\kern -0.18em B}{}\xspace}
\def\Bb      {\ensuremath{\Bbar}\xspace}
 
\def\Bz      {\ensuremath{B^0}\xspace}
\def\Bzb     {\ensuremath{\Bbar^0}\xspace}
\def\BzBzb   {\ensuremath{\Bz {\kern -0.16em \Bzb}}\xspace}
\def\Bu      {\ensuremath{B^+}\xspace}
\def\Bub     {\ensuremath{B^-}\xspace}

\def\BpBm    {\ensuremath{\Bu {\kern -0.16em \Bub}}\xspace}

\mathchardef\Upsilon="7107
\def\Y#1S{\ensuremath{\Upsilon{(#1S)}}\xspace}

\def\FourS {\Y4S}

\mathchardef\Deltares="7101
\mathchardef\Xi="7104
\mathchardef\Lambda="7103
\mathchardef\Sigma="7106
\mathchardef\Omega="710A

\def\Deltabar{\kern 0.25em\overline{\kern -0.25em \Deltares}{}\xspace}
\def\Lbar{\kern 0.2em\overline{\kern -0.2em\Lambda\kern 0.05em}\kern-0.05em{}\xspace}
\def\Sigbar{\kern 0.2em\overline{\kern -0.2em \Sigma}{}\xspace}
\def\Xibar{\kern 0.2em\overline{\kern -0.2em \Xi}{}\xspace}
\def\Obar{\kern 0.2em\overline{\kern -0.2em \Omega}{}\xspace}
\def\Nbar{\kern 0.2em\overline{\kern -0.2em N}{}\xspace}
\def\Xb{\kern 0.2em\overline{\kern -0.2em X}{}\xspace}

\def\mes        {\mbox{$m_{\rm ES}$}\xspace}

\newcommand{\tev}{\ensuremath{\mathrm{\,Te\kern -0.1em V}}\xspace}
\newcommand{\gev}{\ensuremath{\mathrm{\,Ge\kern -0.1em V}}\xspace}
\newcommand{\mev}{\ensuremath{\mathrm{\,Me\kern -0.1em V}}\xspace}
\newcommand{\kev}{\ensuremath{\mathrm{\,ke\kern -0.1em V}}\xspace}
\newcommand{\ev}{\ensuremath{\mathrm{\,e\kern -0.1em V}}\xspace}
\newcommand{\gevc}{\ensuremath{{\mathrm{\,Ge\kern -0.1em V\!/}c}}\xspace}
\newcommand{\mevc}{\ensuremath{{\mathrm{\,Me\kern -0.1em V\!/}c}}\xspace}
\newcommand{\gevcc}{\ensuremath{{\mathrm{\,Ge\kern -0.1em V\!/}c^2}}\xspace}
\newcommand{\mevcc}{\ensuremath{{\mathrm{\,Me\kern -0.1em V\!/}c^2}}\xspace}

\def\mus  {\ensuremath{\rm \,\mus}\xspace}

\def\mus        {\ensuremath{\,\mu{\rm s}}\xspace}    

\def\pep2{PEP-II}

\newcommand{\dedx}{\ensuremath{\mathrm{d}\hspace{-0.1em}E/\mathrm{d}x}\xspace}

\def\gsim{{~\raise.15em\hbox{$>$}\kern-.85em
          \lower.35em\hbox{$\sim$}~}\xspace}
\def\lsim{{~\raise.15em\hbox{$<$}\kern-.85em
          \lower.35em\hbox{$\sim$}~}\xspace}

\xspace

\def\jetset74   {\mbox{\tt Jetset \hspace{-0.5em}7.\hspace{-0.2em}4}\xspace}

\newcommand{\BABARPubYear}    {02}
\newcommand{\BABARPubNumber}  {007}

\newcommand{\SLACPubNumber} {9262}
\newcommand{\LANLNumber} {0000000}

\def\figurebox#1#2#3{
    \def\arg{#3}
    \ifx\arg\empty
    {\hfill\vbox{\hsize#2\hrule\hbox to #2{\vrule\hfill\vbox to #1{\hsize#2\vfill}\vrule}\hrule}\hfill}
    \else
    {\hfill\epsfbox{#3}\hfill}
    \fi}

\begin{document}
\preprint{\babar-PUB-\BABARPubYear/\BABARPubNumber} 
\preprint{SLAC-PUB-\SLACPubNumber} 
\begin{flushleft}
\babar-PUB-\BABARPubYear/\BABARPubNumber\\
SLAC-PUB-\SLACPubNumber\\
hep-ex/\LANLNumber\\[10mm]
\end{flushleft}
\title{
{\large \bf
Rare {\boldmath$B$} Decays into States Containing a {\boldmath$J/\psi$} Meson and  
a Meson with {\boldmath $s\bar s$} Quark Content } 
}


%
\author{B.~Aubert}
\author{D.~Boutigny}
\author{J.-M.~Gaillard}
\author{A.~Hicheur}
\author{Y.~Karyotakis}
\author{J.~P.~Lees}
\author{P.~Robbe}
\author{V.~Tisserand}
\author{A.~Zghiche}
\affiliation{Laboratoire de Physique des Particules, F-74941 Annecy-le-Vieux, France }
\author{A.~Palano}
\author{A.~Pompili}
\affiliation{Universit\`a di Bari, Dipartimento di Fisica and INFN, I-70126 Bari, Italy }
% ...
\collaboration{The \babar\ Collaboration}
\noaffiliation
\date{\today}

\begin{abstract}
We report a study of the $B$ meson decays, $B^+\rightarrow J/\psi \phi K^{+}$, $B^0\rightarrow J/\psi\phi K_{S}^{0}$, $B^0\rightarrow J/\psi\phi$, $B^0\rightarrow J/\psi \eta $ and $B^0\rightarrow J/\psi \eta ^{\prime }$ using {56} million $B\overline B$ events collected at the \FourS\ resonance with the \babar\ detector at the PEP-II $e^+ e^-$ asymmetric-energy storage ring. We measure the branching fractions ${\cal B}(B^+\rightarrow J/\psi \phi K^{+}) = (4.4\pm 1.4(stat) \pm 0.5(syst))\times 10^{-5}$ and ${\cal B}(B^0\rightarrow J/\psi\phi K_{S}^{0}) = (5.1\pm 1.9(stat) \pm 0.5(syst))\times 10^{-5}$, and set upper limits at 90$\%$ confidence level for the branching fractions ${\cal B}(B^0\rightarrow J/\psi\phi) < 9.2\times 10^{-6}$, ${\cal B}(B^0\rightarrow J/\psi \eta )< 2.7\times 10^{-5}$, and ${\cal B}(B^0\rightarrow J/\psi \eta ^{\prime })< 6.3\times 10^{-5}$.
\end{abstract}


\maketitle

\begin{figure}[!htb]
\centerline{\epsfxsize=3.5truein\epsffile{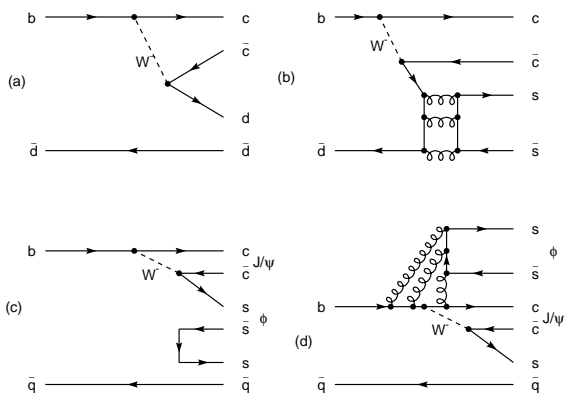}}
\caption{Quark diagrams: (a) tree diagram for $B\rightarrow J/\psi \pi $ and $J/\psi \rho$, (b) rescattering for $B\rightarrow J/\psi \phi$, (c) strange sea quarks and (d) gluon coupling for $B\rightarrow J/\psi \phi K$.}
\label{fig:feyn-jpsiphik}
\end{figure}

Recent observations of the $B$ meson decays $B\rightarrow J/\psi \pi$~\cite{babar-charmonium} and $J/\psi \rho$~\cite{babar-jpsirho} are evidence for the Cabibbo-suppressed transition $b\rightarrow c\overline{c}d$ via the color-suppressed diagram shown in Fig.~\ref{fig:feyn-jpsiphik} (a). Here we present a search for color-suppressed modes with hidden strangeness, $s\overline{s}$, in the final state: $B\rightarrow J/\psi\eta, J/\psi\eta^\prime, J/\psi\phi$ and $J/\psi\phi K$. The decays $B^0\rightarrow J/\psi \eta $ and $B^0\rightarrow J/\psi \eta ^{\prime }$ occur via the same diagram, Fig.~\ref{fig:feyn-jpsiphik} (a), and should have a rate comparable to $B\rightarrow J/\psi \pi $. If large enough samples can be isolated, these $CP$ eigenstates could be used to test $CP$ violation~\cite{beneke}. Models based on the heavy quark factorization approximation by A. Deandrea~\textit{et al.}~\cite{deandrea} are used to predict that the branching fraction for $B^0\rightarrow J/\psi \eta $ is a factor of 4 smaller than that for $B^{0}\rightarrow $ $J/\psi \pi ^{0}$. Assuming that the decay $B^0\rightarrow J/\psi \phi$ is a color-suppressed mode with rescattering as shown in Fig.~\ref{fig:feyn-jpsiphik} (b), then the absence of a signal would indicate that the rescattering effects are negligible. The decay $B\rightarrow J/\psi \phi K$ is a  Cabibbo-allowed and color-suppressed decay via the transition $b\overline{q}\rightarrow c\overline{c}s\overline{s}s\overline{q}$, where the $s\overline{s}$ quark pairs are produced from sea quarks or are connected via gluons as shown in Figs.~\ref{fig:feyn-jpsiphik} (c) and (d), respectively. This particular three-body decay would be of interest in the search for hybrid charmonium states that decay to the final state $J/\psi \phi$~\cite{close}. In this paper we report on branching fractions or upper limits for $J/\psi \eta $, $J/\psi \eta ^{\prime }$, $J/\psi \phi$, $J/\psi \phi K^{+}$, and $J/\psi \phi K_{S}^{0}$.

The data used in this analysis were collected at
the PEP-II asymmetric-energy $e^{+}e^{-}$ storage ring with the $\babar$ detector, fully described elsewhere~\cite{babar-det}. The $\babar$
detector contains a five-layer silicon vertex tracker (SVT) and a forty-layer drift chamber (DCH) in a 1.5-T solenoidal magnetic field. These devices detect charged particles and measure their momentum and
energy loss. Photons and neutral hadrons are detected in a
CsI(Tl) crystal electromagnetic calorimeter (EMC). The EMC detects
photons with energies as low as 20 MeV and identifies electrons by their energy deposition. An internally reflecting ring-imaging
Cherenkov detector (DIRC) of quartz bars is dedicated to charged particle
identification (PID). Penetrating
muons and neutral hadrons are identified by the steel flux return (IFR), which is instrumented with 18-19 layers of resistive plate chambers.

The data correspond to a total integrated luminosity of 
$50.9$ fb$^{-1}$ taken on the $\FourS$
resonance and {6.3} fb$^{-1}$ taken off-resonance at an energy
0.04 GeV below the $\FourS$ mass and
below the threshold for $B\overline{B}$ production. In this sample, there are $55.5\pm0.6$ million $B\overline{B}$ events ($N_{B
\overline{B}}$).

In this analysis, all charged track candidates are required to have at least 12 DCH hits and transverse momentum greater than 100 MeV/$c$. The track candidates not associated with a $K^0_{S}$ decay must also
originate near the nominal beam spot. The muon, electron, and kaon candidates must have a polar angle in
radians of $0.3<\theta _{\mu }<2.7$, $0.410<\theta _{e}<2.409$, and $
0.45<\theta _{K}<2.50$, respectively.
In addition, all charged kaon candidates are required
to have a laboratory momentum greater than 250 MeV/$c$.
These requirements ensure the selection of tracks in the regions where the acceptance is well understood by the PID systems.

Photon candidates are identified from energy deposited in contiguous EMC crystals, summed together to form a cluster with total energy greater than 30 MeV and a shower shape consistent with that expected for 
electromagnetic showers. 

Electron candidates are required to have a good match 
between the expected and measured energy loss ($\dedx$) in the DCH, and 
between the expected and measured Cherenkov angle in the DIRC. 
The measurements of 
the ratio of EMC
shower energy to DCH momentum, and the number of EMC crystals associated with the track
candidate must be appropriate for an electron. 

Muons are selected based on the energy deposited in the EMC, the number and distribution of hits in the IFR,  the match between the IFR hits and the extrapolation of the DCH track into the IFR, and the depth of penetration of the track into the IFR.

Charged kaon and pion candidates are selected based on energy loss
information from the SVT and DCH and the Cherenkov angle measured by the DIRC.

The intermediate states in the indicated decay modes used in this analysis, $J/\psi \left(
ee,\mu \mu \right) $, $\phi \left( K^{+}K^{-}\right) $, $\eta(\gamma\gamma, \pi^{+}\pi ^{-}$$\pi^{0})$, $\eta ^{\prime }\left( \eta
\left( \gamma \gamma \right) \pi ^{+}\pi ^{-}\right) $, $\pi ^{0}\left(
\gamma \gamma \right)$, and $K_{S}^{0}\left( \pi ^{+}\pi ^{-}\right) $, are selected with the mass intervals in Table~\ref{table-mass}. 
Since $B^{0}\rightarrow J/\psi \eta$ and $B^{0}\rightarrow J/\psi \eta^{\prime}$ involve decays of a pseudoscalar meson into a vector and a pseudoscalar meson, 
the angular distribution is proportional to $\sin ^{2}\theta_{\ell}$, where $\theta_{\ell}$ is the helicity 
angle~\footnote{
In the reaction, $Z\rightarrow X+Y, X\rightarrow a+b$, the helicity angle of particle $a$ is defined as the angle measured in the particle $X$ rest frame between the direction of particle $a$ and the direction opposite to particle $Z$.} 
of the lepton from the $J/\psi$. Hence an additional requirement of $
\left| \cos \theta _{\ell}\right| <0.8$ is applied to reject continuum and other backgrounds. The $\eta$ candidates are rejected
if either of the associated photons,
in combination with any other photon in the event,
forms
a $\gamma \gamma $ mass within 20 MeV/$c^{2}$ of the $\pi^0$ mass. For the mode $B^{0}\rightarrow J/\psi \eta(\gamma\gamma)$, the $\eta$ candidate is required
to have
$\left| \cos \theta _{\gamma }^{\eta }\right| <0.8$, where $
\theta _{\gamma }^{\eta }$ is the photon helicity angle in the $\eta$ rest
frame. 
This rejects combinatoric background due to random pairs of photons
that typically have a  photon helicity angle that
peaks at 0 or 180 degrees. 
For the $\eta ^{\prime }\rightarrow \eta \left( \gamma \gamma \right) \pi
^{+}\pi ^{-}$ candidates, we use the same $\eta$ selection criteria for the $\eta$ described above, including the
$\pi^0$ veto. 

An additional requirement is applied to separate two-jet continuum events from the
more spherical $B$ meson decays. The angle $\theta _{T}$ between the thrust direction of the $B$
meson candidate and the thrust direction of the remaining tracks in the
event is calculated. We require $\left| \cos \theta _{T}\right| <0.8$, 
since these thrust axes are uncorrelated and
the distribution 
in $\cos\theta _{T}$ 
is flat for $B\overline{B}$ events, 
while the distribution is peaked at $\cos \theta _{T}=\pm 1$ for continuum events. 

\begin{table}[!htb]
\caption{Mass regions for selection of intermediate particles.}
\begin{center}
\begin{tabular}{lrcccl}
\hline\hline
Mode & \multicolumn{5}{c}{Mass Range (GeV/$c^{2}$)} \\ \hline
$J/\psi \rightarrow e^+e^-$ & $2.95$&$<$&$M(e^{+}e^{-})$&$<$&$3.14$ \\ 
$J/\psi \rightarrow \mu^+ \mu^- $ & $3.06$&$<$&$M\left( \mu ^{+}\mu ^{-}\right)$&$ <$&$3.14$
\\ 
$\phi \rightarrow K^{+}K^{-}$ & $1.004$&$<$&$M\left( K^{+}K^{-}\right) $&$<$&$1.034$ \\ 
$K_{S}^{0}\rightarrow \pi ^{+}\pi ^{-}$ & $0.489$&$<$&$M\left( \pi ^{+}\pi
^{-}\right)$&$ <$&$0.507$ \\ 
$\eta \rightarrow \gamma \gamma $ & $0.529$&$<$&$M\left( \gamma \gamma \right)
$&$<$&$0.565 $ \\ 
$\eta \rightarrow \pi ^{+}\pi ^{-}\pi ^{0}$ & $0.529$&$<$&$M\left( \pi ^{+}\pi
^{-}\pi ^{0}\right) $&$<$&$0.565$ \\ 
$\eta ^{\prime }\rightarrow \eta \pi ^{+}\pi ^{-}$ & $0.938$&$<$&$M\left( \eta \pi
^{+}\pi ^{-}\right)$&$ <$&$0.978$ \\ 
$\pi ^{0}\rightarrow \gamma \gamma $ & $0.120$&$<$&$M\left( \gamma \gamma \right)
$&$<$&$0.150$ \\ \hline
\end{tabular}
\end{center}
\label{table-mass}
\end{table}

The intermediate candidates are combined to construct the $B$ candidates for the six decay modes under study. The estimation of the signal and the background employs two kinematic variables: the energy difference $\Delta E$ between the energy of the $B$ candidate and the beam energy $E_{b}^{*}$ in the $\FourS$ rest frame; and the energy-substituted mass $\mes =\sqrt{\left( E_{b}^{*}\right)^2 -\left( P_{B}^{*}\right) ^{2}}$, where 
$ P_{B}^{*}$ is the reconstructed momentum of the $B$ candidate in the $\FourS$ frame.
Typically these two weakly correlated variables form a two-dimensional
Gaussian distribution for the $B$ meson signal but not for background. The resolutions in $\Delta E$ and $\mes$ are decay mode dependent. 
A signal region for each mode is defined as a rectangular region in the 
$\Delta E$ versus $\mes$ plane, listed
in Table~\ref{table-upperlimit-bf}. The $\mes$ range is given in term of $\mes-m_{B}$, where $m_{B}$ is the mass of $B$ meson. The number of data events, $n_0$, observed in the
signal region for each mode is listed in Table~\ref{table-upperlimit-bf}. 

The efficiencies for each mode are determined by Monte Carlo simulation.
The simulations of $J/\psi \phi K$ and $J/\psi \phi$ decays assumed three- and two-body phase space, respectively, with
unpolarized $J/\psi$ and $\phi$ decays.
The $J/\psi \eta$ and  $J/\psi \eta^{\prime }$ simulations used 
the angular correlations determined by the helicity amplitude.

The backgrounds in the $\mes$ distribution are composed of two components: a combinatoric background, whose shape is described by an ARGUS function~\cite{argus-function}, and a peaking background that peaks in the signal region and is described by a Gaussian function. The sources of combinatoric background are the continuum events and two categories of $B\overline{B}$ events: decays with a leptonic $J/\psi$ decay, and those without. 
Monte Carlo simulation studies show that the source of the peaking background is $B\overline{B}$ events that contain a leptonic $J/\psi$ decay.

The shape of the ARGUS function is determined mode by mode by fitting to the $\mes$ distribution of candidates in an enhanced fake $J/\psi$ sample, which is obtained by reversing the normal lepton identification requirements. 
 
The normalization of the combinatoric background for each mode is obtained from a fit to the $\mes$ distributions in the $\Delta E$ signal region of the on-peak data. The integral of the ARGUS function in the signal region is $n_{C}$, the number of combinatoric background events.

The peaking background is determined from a fit to the $\mes$ distribution of Monte Carlo $\B\Bb$ events with leptonic $J/\psi$ decays using the sum of a Gaussian and an ARGUS function. The number of peaking background events $n_{P}$ is the integral of the Gaussian function in the signal region.

The total number of background events ($n_{b}$) and the uncertainty on this number ($\sigma_b$) are calculated from the fit value of $n_{C}$ and $n_{P}$ and their errors. The values of $n_{b}$ and $\sigma_b$ are listed in Table~\ref{table-upperlimit-bf} for all modes.
The combinatoric background is by far the dominant background in all modes except the $B^{0}\rightarrow J/\psi \eta (\pi^+\pi^-\pi^0)$ mode, where the peaking component is $\sim20\%$ of the total background.

In Table~\ref{table-systematics}, we list the contribution to the systematic error from the uncertainty 
on each of the following quantities: $N_{B\overline B}$; 
secondary branching fractions~\cite{pdg}; Monte Carlo statistics; PID, tracking, and photon detection efficiencies, which are based on the study of control samples; and background parameterization, which is estimated using $\Delta E$ sideband information. 

Additional systematic uncertainties due to the decay model dependence are 
estimated for the modes $J/\psi \phi$, $J/\psi \phi K^{+}$, and $J/\psi \phi K_{S}^{0}$. Monte Carlo simulations are used to determine how much  the efficiency depends on assumptions about intermediate resonances and angular distributions. Two samples are generated for each of the three modes with decay distributions determined by the assumed polarization of the vector daughter mesons, rather than by phase space. One sample is  generated with $100\%$ transversely polarized $J/\psi$ and $\phi$ mesons, and the other with $100\%$ longitudinally polarized $J/\psi$ and $\phi$ mesons. The resulting relative change in efficiency is entered as a fractional systematic 
error in Table~\ref{table-systematics}. An additional check based on Monte Carlo samples with an intermediate state gives negligible effect. 

The total systematic error for each mode combines all these separate errors in quadrature
and is listed (Total) in Table~\ref{table-systematics}. 

\begin{figure}[!htb]
\begin{center}
\includegraphics[height=7cm]{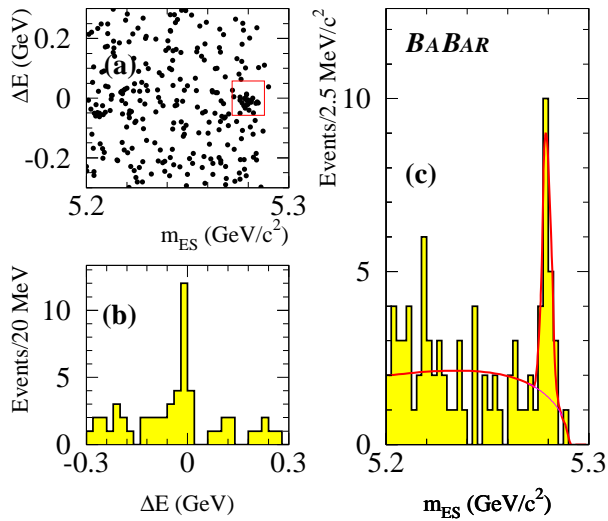}
\caption{The $\Delta E$ and $\mes$ distributions for $B^{+}\rightarrow J/\psi \phi K^+$. 
The $\Delta E$ vs. $\mes$ event distribution is shown in (a) with a 
small rectangle corresponding to the signal region selection defined
in Table II. 
The $\Delta E$ projection with a $\mes$ signal region selection is shown in (b).
The $\mes$ projection with a $\Delta E$ signal region selection is shown in (c).
The solid line in (c) is the fit
described in the text. 
The Gaussian component
includes both the signal and peaking background.}
\label{fig:jpkub}
\end{center}
\end{figure}

\begin{figure}[!htb]
\begin{center}
\includegraphics[height=7cm]{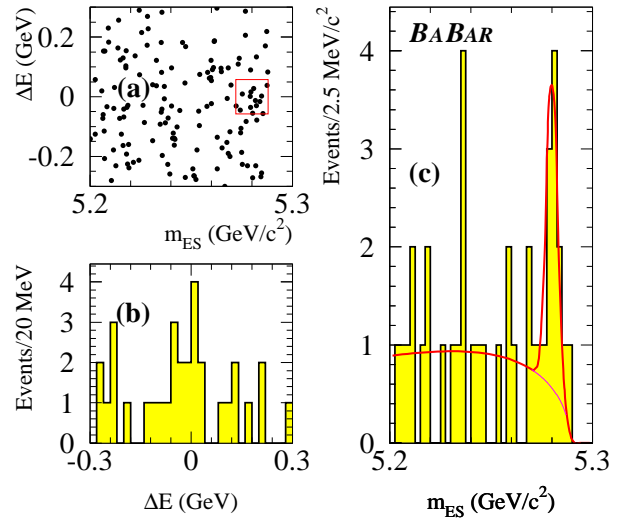}
\caption{The $\Delta E$ and $\mes$ distributions for $B^{0}\rightarrow J/\psi \phi K^0_S$. 
The descriptions of Figs. 3(a), (b) and (c) follow those of Figs. 2(a), (b) and (c), respectively.}
\label{fig:jpksub}
\end{center}
\end{figure}

There is evidence for signals in the $J/\psi \phi K^{+}$ and $J/\psi \phi K_{S}^{0}$ modes. 
The results are shown in  Figs.~\ref{fig:jpkub} and \ref{fig:jpksub}. 
The Poisson probability that the background $n_b$ fluctuates up to the observed number of events, $n_0$, or higher 
is $7.7\times10^{-6}$ for $J/\psi \phi K^{+}$ and $4.2\times10^{-5}$ for $J/\psi \phi K_{S}^{0}$. 
The branching fraction for these modes is determined by a simple subtraction of events in the signal region that 
yields the number of signal events, $n_{s}=n_{0}-n_b$. 
The calculation of the branching fraction is based on
the efficiency, $n_{s}$, $N_{B\overline{B}}$, and the secondary branching fractions for the $J/\psi ,$ $\phi $, and $
K_{S}^{0}$ from Ref.~\cite{pdg}. 
The results are summarized in Table~\ref{table-upperlimit-bf} where
the first error is the statistical error and the second is the systematic error, listed in Table~\ref{table-systematics}. The derived result for $B^0 \rightarrow J/\psi \phi K^0$ is also shown in Table~\ref{table-upperlimit-bf}.

For modes with no signal or limited statistical evidence
($J/\psi \phi $, $J/\psi \eta $, $J/\psi \eta ^{\prime }$), 
we determine both a central confidence interval and an upper
limit interpretation for the branching fraction.
The upper limit method uses $n_{0}$, $n_{b}$, and $\sigma_b$, in the signal region, and the total systematic uncertainty $\sigma _{T}$. Assuming the two uncertainties ($\sigma _{b},\sigma _{T}$) are uncorrelated and Gaussian, 
the Bayesian upper limit on the number of events ($N_{90\%}$) is obtained by
folding the Poisson distribution with two normal distributions for
these two uncertainties and integrating it to the $90\%$ confidence level (C.L.). In Table~\ref{table-upperlimit-bf} we list for each mode the efficiency, the  number of observed events, the expected number of background events, the $90\%$ C.L. upper limit for observed events, 
the corresponding branching fraction limit and a central interval for the branching fraction. 
The upper limit obtained from the combination of the two $B^0\rightarrow J/\psi \eta$ modes is shown in Table~\ref{table-upperlimit-bf}. The mean value of the branching fraction is calculated for $B^0\rightarrow J/\psi \phi$ and $B^0\rightarrow J/\psi \eta ^{\prime }$. We also combine the observed numbers of events for the two $B^0\rightarrow J/\psi \eta$ modes to calculate a branching fraction of $(1.6\pm0.6(stat.)\pm0.1(syst.))\times 10^{-5}$. The Poisson probability that the background fluctuates up to the observed number of events or higher is $2.5\times10^{-5}$ for the combined result.

\begin{table*}
\caption{Branching fractions and 90\% C.L. upper limits.}
\begin{center}
\footnotesize{
\begin{tabular}{lcccccccc}
\hline\hline\\[-0.2cm]

Mode &\multicolumn{2}{c}{Signal Region}& Efficiency& $n_{0}$ & $n_{b}\pm \sigma _{b}$ & $N_{90\%}$ &~~~$90\%$ C.L. Upper Limit & ~~~Branching Fraction \\ 
&$\Delta E$ (MeV)&~~~$|\mes-m_{B}|$ (MeV/$c^{2}$)&&&&&$(10^{-5})$ &~~~ $(10^{-5})$ \\\hline \\[-0.2cm]

$J/\psi \phi K^{+}$ &57.0&~~~8.0& $10.6\%$&23 & $7.8\pm 0.6$ &&&~~~$
4.4\pm 1.4\pm 0.5$ \\ 
$J/\psi \phi K_{S}^{0}$ &57.0&~~~8.0& ${8.6\%}$&13 & {${3.3\pm 0.4}$} &&& 
~~~$5.1\pm 1.9\pm 0.5$ \\ 
$J/\psi \phi K^{0}$ &&&&&&&&~~~ $10.2\pm 3.8\pm 1.0$\\ 

$J/\psi \phi $ &57.0&~~~8.0&${12.1\%}$& ${1}$ & ${0.3\pm 0.2}$ & ${3.60}$ & $~~~<0.9$&~~~${0.18\pm0.26\pm0.03}$\\
$J/\psi \eta ^{\prime }$&100.0&~~~10.0& ${2.5\%}$& ${0}$ &${0.5\pm 0.3}$ & {${1.81}$} &$~~~<6.3$&~~~${-1.7\pm1.0\pm0.2}$ \\ 
$J/\psi \eta \left( \gamma \gamma \right) $&100.0&~~~10.0 &{15.5\%} &${8}$ & ${1.7\pm 0.4}$ & {${11.5}$} & $~~~<2.9$&\\
$J/\psi \eta \left( \pi^+\pi^-\pi^0 \right) $&72.0&~~~10.0&${8.7\%}$ & ${4}$ & ${1.5\pm 0.9}$ & {${6.76}$}& $~~~<5.1$&\\ 
$J/\psi \eta$ combined & &&&&&& $~~~<2.7$&~~~$1.6\pm 0.6\pm 0.1$ \\ \hline
\end{tabular}
}
\end{center}
\label{table-upperlimit-bf}
\end{table*}

\begin{table*}[!htb]
\caption{Systematic error summary on the branching fractions. All are fractional uncertainties in percent.}
\begin{center}
\footnotesize{
\begin{tabular}{lccccccc}
\hline\hline\\[-0.2cm]
Mode &$N_{B\overline B}$ & Secondary & Monte Carlo & PID, Tracking,& Background&Model & Total \\
& & Branching Fractions&Statistics&Photon Detection& Parameterization&& \\\hline\\[-0.2cm]
$J/\psi \phi K^+$ &1.1& 2.2 & 1.6 & 8.2 & 5.9 &0.4& 10.4 \\ 
$J/\psi \phi K_{S}^0$ &1.1& 2.2 & 2.1 & 8.3 & 1.9&0.9 & 9.3 \\ 
$J/\psi \phi $ &1.1& 2.2 & 1.6 & 6.7& 12.0&1.0 & 14.1 \\ 
$J/\psi \eta ^{\prime }$&1.1 & 3.8 & 4.6 & 9.3 & 7.1&- & 13.3 \\
$J/\psi \eta \left( \gamma \gamma \right) $&1.1 & 1.8 & 1.6 & 6.0 & 6.9 &-& 9.5 \\ 
$J/\psi \eta \left(\pi^+\pi^-\pi^0 \right) $&1.1 & 2.4 & 2.2 & 7.7 & 8.0&-
& 11.6 \\ \hline
\end{tabular}
}
\end{center}
\label{table-systematics}
\end{table*}

In summary, we determine the branching fraction of $B\rightarrow$ $J/\psi \phi K$ in two
modes, ${\cal B}$($B^+\rightarrow J/\psi \phi K^{+}$)=
{${(4.4\pm 1.4\pm 0.5)\times 10}^{-5}$} and ${\cal B}$($B^0\rightarrow J/\psi
\phi K_{S}^{0}$)={$(5.1\pm 1.9\pm 0.5)\times 10^{-5}$}. The branching fraction of $
B\rightarrow J/\psi \phi K$ is consistent with a CLEO~\cite{cleo} result, $\left(8.8_{-3.0}^{+3.5}\pm 1.3\right) \times 10^{-5}$. 
Upper
limits have been determined for the modes $B^0\rightarrow J/\psi \phi ,$ $
J/\psi \eta $, and $J/\psi \eta ^{\prime }$. The upper limit on $B^0\rightarrow J/\psi \eta$ is a significant improvement over the previous best limit of $<1.2\times 10^{-3}$ at $90\%$ C.L., from the L3 Collaboration~\cite{L3}. The combined branching fraction for $B^0\rightarrow J/\psi \eta$ is comparable to the $
B^0\rightarrow J/\psi \pi ^{0}$ branching fraction~\cite{babar-charmonium}. 
The search and resulting upper limits on the branching fractions for 
$B^0\rightarrow J/\psi \eta ^{\prime }$ and $B^0\rightarrow J/\psi \phi$ 
are presented.

\label{sec:Acknowledgments}

We are grateful for the excellent luminosity and machine conditions
provided by our \pep2\ colleagues, 
and for the substantial dedicated effort from
the computing organizations that support \babar.
The collaborating institutions wish to thank 
SLAC for its support and kind hospitality. 
This work is supported by
DOE
and NSF (USA),
NSERC (Canada),
IHEP (China),
CEA and
CNRS-IN2P3
(France),
BMBF and DFG
(Germany),
INFN (Italy),
NFR (Norway),
MIST (Russia), and
PPARC (United Kingdom). 
Individuals have received support from the 
A.~P.~Sloan Foundation, 
Research Corporation,
and Alexander von Humboldt Foundation.
$\smallskip $


\begin{thebibliography}{99}
\bibitem{babar-charmonium}$\babar$ Collaboration, B. Aubert $et\ al.$, Phys. Rev. \textbf{D65}, 32001 (2002).

\bibitem{babar-jpsirho}CLEO Collaboration, M. Bishai $et\ al.$, Phys. Lett. \textbf{B369}, 186 (1996); $\babar$ Collaboration, hep-ex/0209013, submitted to Phys. Rev. Lett.


\bibitem{beneke} M. Beneke, G. Buchalla, and I. Dunietz, Phys. Lett. \textbf{B393}, 132 (1997).
   
\bibitem{deandrea}A. Deandrea \textit{et al.}, Phys. Lett. \textbf{B318}, 549 (1993).

\bibitem{close}F. E. Close $et\ al.$, Phys. Rev. \textbf{D57}, 5653 (1998).


\bibitem{babar-det}$\babar$ Collaboration, B. Aubert $et\ al.$, Nucl. Instr. and Methods \textbf{A479}, 1 (2002).

\bibitem{argus-function}ARGUS Collaboration, H. Albrecht \textit{et al.}, Z. Phys \textbf{C48}, 543 (1990).

\bibitem{pdg}Particle Data Group, D.E. Groom \textit{et al.}, Eur. Phys. J. C \textbf{15}, 1 (2000).

\bibitem{L3}L3 Collaboration, M. Acciarri $et\ al.$, Phys. Lett. \textbf{B391}, 481 (1997).

\bibitem{cleo}CLEO Collaboration, A. Anastassov $et\ al.$, Phys. Rev. Lett. \textbf{84}, 1393 (2000).


\end{thebibliography}
\end{document}